\documentclass[aps,prl,twocolumn,superscriptaddress,preprintnumbers]{revtex4}
\usepackage{graphicx}
\usepackage{amsmath,amssymb,bm}
\usepackage[mathscr]{eucal}
\usepackage[colorlinks,linkcolor=blue,citecolor=blue]{hyperref}

\newif\ifarxiv
\arxivfalse

\begin		{document}
\def\st    {\begin{equation}}
\def\stp    {\end{equation}}

\def\Arxiv      #1 [#2]{\href{http://arxiv.org/abs/#1}{{\tt arXiv:#1 [#2]}}\,}

\definecolor{Blueberry}{rgb}{0.25,0,0.65}
\definecolor{Strawberry}{rgb}{0.65,0,0.25}

\title
    {
    Holographic turbulence 
    }
\author{Allan~Adams}
\author{Paul~M.~Chesler}`
\author{Hong~Liu}

\affiliation
    {%
Department of Physics, 
MIT, 
Cambridge, MA 02139, USA 
    }%

\date{\today}

\begin{abstract}
We construct turbulent black holes in asymptotically AdS$_4$ spacetime by numerically solving Einstein equations.
Both the dual holographic fluid and bulk geometry display signatures of an inverse cascade with the bulk geometry being well approximated by 
the fluid/gravity gradient expansion.  We argue that statistically steady-state black holes dual to $d$ dimensional turbulent flows
have horizons which are approximately fractal with fractal dimension $D=d+4/3\,$. 
\end{abstract}

\preprint{MIT-CTP-4460}


\maketitle

\textit{Introduction}.---According to holography, turbulent flows in relativistic boundary
conformal field theories should be dual to dynamical black hole solutions in  
asymptotically AdS$_{d+2}$ spacetime with $d$ the number of spatial dimensions the turbulent flow lives in.
This immediately raises many interesting questions about gravitational dynamics.
For example, what distinguishes turbulent black holes from 
non-turbulent ones? What is the gravitational origin 
of energy cascades and the Kolmogorov scaling observed in turbulent fluid flows? 

Gravitational dynamics can also provide insight into
turbulence itself. This is particularly relevant for problems
where  physics beyond hydrodynamics may play a crucial role
in turbulent evolution, since holography provides a complete
description of physics valid on all length
scales. For superfluid turbulence, where vortex dynamics
and turbulent evolution are not governed by hydrodynamics, holography has
already yielded valuable insight and demonstrated two
dimensional holographic superfluids can exhibit a direct
energy cascade to the UV \cite{Adams:2012pj}. Likewise, having control of regimes beyond
the hydrodynamic description of turbulence of normal fluids
may allow one to study the domain of validity and the
late-time regularity of solutions to the Navier-Stokes equation.

In this paper we take a first step towards studying holographic turbulence
by numerically constructing black hole solutions in asymptotically AdS$_4$ spacetime
dual to $d = 2$ turbulent flows, where energy flows from the UV to the IR in an inverse 
cascade.  We 
propose a simple geometric measure to distinguish turbulent black 
holes from non-turbulent ones: 
the horizon of a turbulent black hole exhibits a 
fractal-like structure with effective fractal dimension $D = d + 4/3$.  The $4/3$ in this formula can be understood
as the geometric counterpart of the rapid entropy growth implied by the
Kolmogorov scaling.

\textit{Numerics and Gravitational Description}.---We generate turbulent evolution by numerically solving Einstein's equations  
\begin{equation}
\label{eq:ein}
R_{MN} - {\textstyle \frac{1}{2}} G_{MN} \left ( R - 2 \Lambda \right ) = 0,
\end{equation}
with cosmological constant $\Lambda = -3$.  Our numerical scheme for solving Einstein's equations
is outlined in \cite{Chesler:2010bi, Chesler:2008hg, Cardoso:2012qm} and will be further elaborated on in a coming
paper \cite{CY}.  In what follows we focus on some of the salient details.

We employ a characteristic formulation of Einstein's equations 
and choose the metric ansatz
\begin{equation}
\label{eq:metric}
ds^2 = r^2 g_{\mu \nu}(x,r)dx^\mu dx^\nu + 2 dr dt,
\end{equation}
with Greek indices $(\mu,\nu)$ running over the boundary spacetime coordinates.  
The coordinate $r$ is the AdS radial coordinate with $r = \infty$ corresponding the the AdS boundary.  
The metric ansatz (\ref{eq:metric}) is invariant under the residual diffeomorphism 
$r \to r + \xi(x)$ for arbitrary $\xi(x)$.  Since the geometry we study contains a black brane with planar topology,
we fix the residual diffeomorphism invariance by demanding the apparent horizon 
be at $r = 1$.  Horizon excision is then performed by restricting the computational domain to $r \geq 1$.

The AdS boundary is causal and therefore boundary conditions must be imposed there.
Solving Einstein's equations with a series expansion about $r = \infty$, one finds an asymptotic expansion of the form 
$g_{\mu \nu}(x,r) = g^{(0)}_{\mu \nu}(x) + \dots + g_{\mu \nu}^{(3)}(x)/ r^3 + \dots$.
All omitted terms in the  expansion 
are determined by $g^{(0)}_{\mu \nu}(x)$ and $g^{(3)}_{\mu \nu}(x)$ and their derivatives.  The expansion coefficient 
$g^{(0)}_{\mu \nu}(x)$ corresponds to the metric the dual turbulent flow lives in.  Hence we choose
the boundary condition $\lim_{r \to \infty} g_{\mu \nu}(x,r) = \eta_{\mu \nu}$.   The expansion coefficient $g_{\mu \nu}^{(3)}(x)$
is determined by solving Einstein's equations and  
encodes the expectation value of the boundary stress tensor \cite{deHaro:2000xn}
\begin{align}
\label{eq:bdstress}
 \langle T_{\mu \nu}(x)\rangle = \frac{3 }{16 \pi G_{\rm N}}  \left [g_{\mu \nu}^{(3)}(x) +{\textstyle \frac{1}{3}}\eta_{\mu \nu} g_{00}^{(3)}(x) \right ],
\end{align}
where $G_{\rm N}$ is Newton's constant.

We choose initial data corresponding to a locally boosted black brane with metric
\footnote
  {
  In the characteristic formulation of Einstein's equations sufficient initial data 
  consists of the conserved boundary 
  densities $\langle T^{0 \mu}\rangle$ and the rescaled spatial metric 
  $\hat g_{ij}\equiv  g_{ij}/\sqrt{{\rm det}g_{ij}}$ \cite{Cardoso:2012qm,CY}.  
  With these quantities specified at $t = 0$ all other components of the metric 
  can be determined by solving Einstein's equations \cite{Cardoso:2012qm,CY}.
  Therefore, when using (\ref{eq:gradmetric}) as initial data we only use $\hat g_{ij}$
  and $\langle T^{0 \mu}\rangle$.
  }
\begin{equation}
\label{eq:gradmetric}
g_{\mu \nu} = (\mathcal R/r)^2 \left [ \eta_{\mu \nu} + \left (r_h/\mathcal R \right )^3 u_\mu u_\nu  \right ],
\end{equation}
where $u_\mu(x)$ is the local boost velocity and $r_{ h}(x) = 4 \pi T(x)/3$ with $T(x)$ the 
local temperature of the brane.  The function $\mathcal R$ satisfies 
$\partial \mathcal R/\partial r =[1 {+} {\textstyle \left ( r_h /\mathcal R \right )^3} \bm u^2 ]^{1/2}.$

We work in a periodic spatial box of size $\Delta x$.   We choose the initial boost velocity 
\begin{equation}
\label{eq:initialvelocity}
u_x(x,y) = \delta u_x(x,y), \ \ u_y(x,y) = \cos Q x + \delta u_y(x,y),
\end{equation}
with $Q= 20 \pi/\Delta x$.  The small fluctuations $\delta u_i$ are present to break the symmetry of the initial 
conditions.  We choose $\delta u_i$ to be a sum of the first four spatial Fourier modes with random coefficients
and phases and adjust the 
overall amplitude of $\delta u_i$ such that $|\delta u_i| < 1/5$. 
These initial conditions are unstable and capable of producing subsequent turbulent evolution
if the Reynolds number $Re$ is sufficiently large. For our initial conditions $Re \sim T \Delta x$.  
We choose box size $\Delta x = 1500$ and the initial temperature $4 \pi T/3 = 1$.

We discretize Einstein's equations using pseudospectral methods and represent the radial dependence of all 
functions in terms of an expansion of $20$ Chebyshev polynomials and the $x-y$ dependence 
of all functions in terms of an expansion of $305$ plane waves.  We then evolve the discretized geometry 
for 3001 units of time.

\begin{figure}[ht!]
\includegraphics[scale = 0.37]{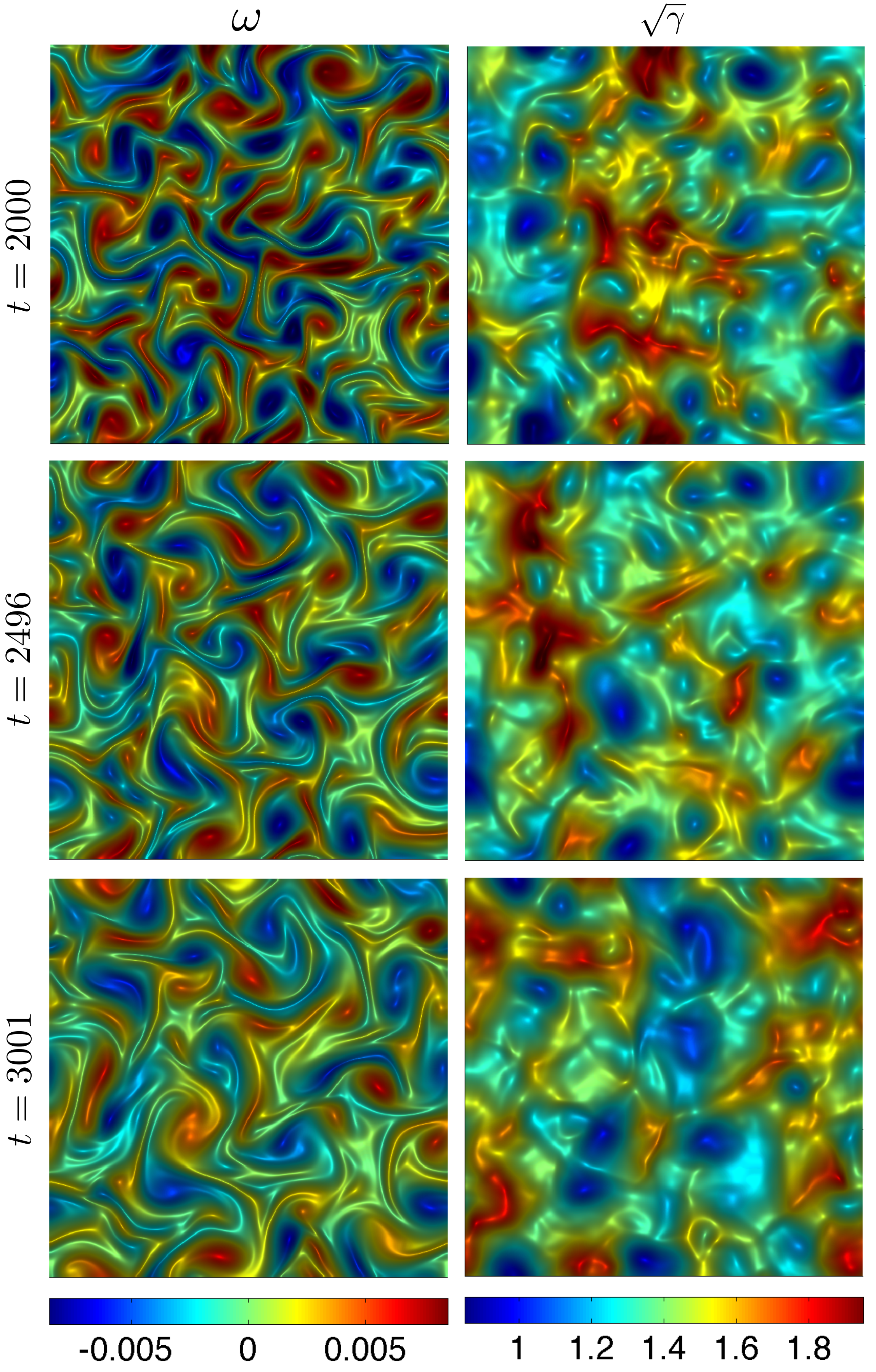}
\caption{Left: The boundary vorticity $\omega$ at three times.   Right: the horizon area element $\sqrt{\gamma}$ at the same three times.
\label{fig:vortictyandarea}
} 
\end{figure}

\textit{Results and Discussion}.---To illustrate the turbulent flow generated by solving Einstein's equations
in Fig.~\ref{fig:vortictyandarea} we plot of the boundary vorticity
$\omega \equiv \varepsilon^{\mu \nu \alpha} u_\mu \partial_\nu u_{\alpha}$
at three different times. 
To compute the vorticity we first extract the boundary stress tensor $\langle T^{\mu \nu} \rangle$ from the metric via Eq.~(\ref{eq:bdstress}).  
We then define the fluid velocity $u^\mu$ as the normalized $(u_\mu u^\mu = -1)$
future-directed time-like eigenvector of $\langle T^{\mu}_{\ \nu} \rangle$,
\begin{equation}
\label{eq:fluidvel}
\langle T^{\mu}_{\ \nu} \rangle u^\nu = - \epsilon u^\mu,
\end{equation}
with $\epsilon$ the proper energy density.

During times $t = 0$ through $t \sim 700$ our system experiences 
an instability which drives our initial state into turbulent evolution.
At all times shown in Fig.~\ref{fig:vortictyandarea} there is little sign of the initial 
sinusoidal structure present in the initial data (\ref{eq:initialvelocity}).
At time $t = 2000$ there are many vortices present
with fluid rotating clockwise (red) and counterclockwise (blue).
During the latter evolution seen at times $t = 2496$ and $3001$ 
isolated vortices with the same rotation tend to merge together to produce larger and larger vortices.
The merging of vortices of like-rotation to produce larger vortices is a tell tale signature of an inverse cascade.

\begin{figure}[h!]
\includegraphics[scale = 0.27]{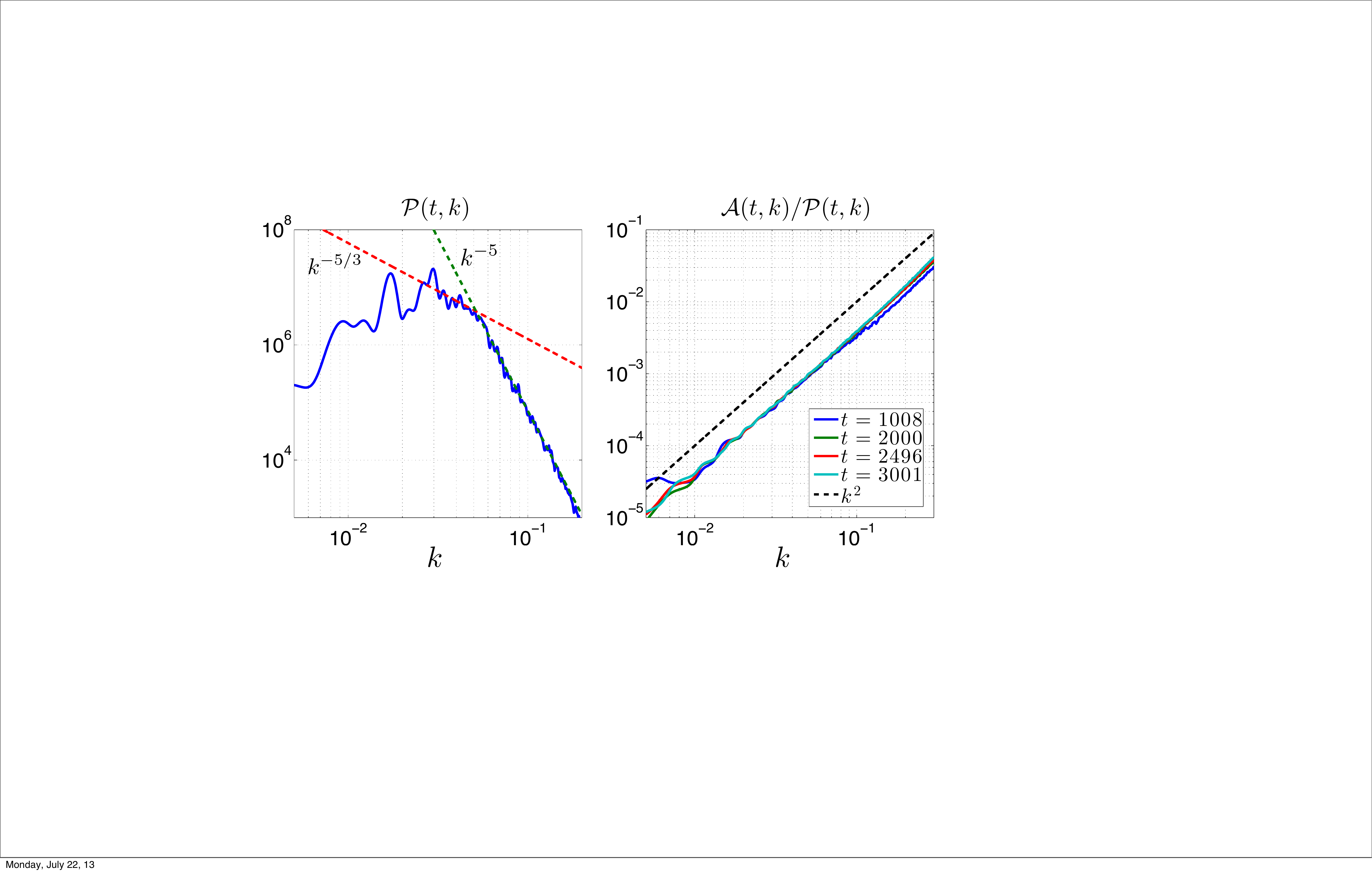}
\caption{Left: The velocity power spectrum $\mathcal P$ at time $t = 1008$. 
Right:  The normalized horizon curvature power spectrum $\mathcal A/\mathcal P$
at four different times.
\label{fig:spectrum}
} 
\end{figure}

It is interesting to compare our results to the Kolmogorov theory of 
turbulence.  A classic result  from Kolmogorov's  theory  
is that for driven steady-state turbulence the power spectrum $\mathcal P$
of the fluid velocity,
\begin{equation}
\mathcal P(t,k) \equiv \frac{\partial}{\partial k} \int\limits_{|\bm k'| \leq k}  \frac{d^d k'}{(2 \pi)^d} | \tilde {\bm u}(t,\bm k')|^2,
\end{equation}
with $\tilde {\bm u}(t,\bm k) \equiv \int d^d x \, \bm u(t,\bm x)\, e^{-i \bm k \cdot \bm x}$, obeys the scaling 
\begin{equation}
\label{eq:kol}
\mathcal P(t,k) \sim k^{-5/3},
\end{equation}  
in an \textit{inertial range} $k \in (\Lambda_-,\Lambda_+)$.  Despite the fact that 
our system is not driven or in a steady-state configuration we do see hints of the Kolmogorov scaling.
In Fig.~\ref{fig:spectrum} we plot $\mathcal P$ at time $t = 1008$.  
Our numerical results are consistent with the scaling (\ref{eq:kol}) in the 
inertial range $k \in (0.025,0.055)$.   As we are not driving the system, 
evidence of the $k^{-5/3}$ scaling is transient and destroyed
first in the UV, with the UV knee at $k = 0.055$ shifting to the IR as time progresses.
Beyond the inertial range the spectrum decreases like $\mathcal P \sim k^{-p}$ 
with $p \sim 5$ until $k \sim 0.15$.  We comment further below on the UV behavior of $\mathcal P$.

The  inverse cascade also manifests itself in gravitational quantities.  One interesting quantity to consider is the horizon
area element $\sqrt{\gamma}$.  In our coordinate system and in the limit of large Reynolds number $Re \gg 1$
the event and apparent horizons approximately coincide at $r = 1$
and the horizon area element is $\sqrt{\gamma} \approx \sqrt{-g}|_{r = 1}$
\footnote
  {
  In the fluid/gravity gradient expansion
  the apparent and event horizons coincide up to second order in gradients.  Hence their
  positions should coincide in the $Re \to \infty$ limit.
  }.
Also included in Fig.~\ref{fig:vortictyandarea} are plots of $\sqrt{\gamma}$.  
At $t = 2000$ $\sqrt{\gamma}$ exhibits structure over a large hierarchy of scales
and is fractal-like in appearance.  We comment more on this further below.
However, as time progresses $\sqrt{\gamma}$ becomes smoother and 
smoother just as the fluid vorticity $\omega$ does due to the inverse cascade.

The velocity power spectrum $\mathcal P$ also imprints 
itself in bulk quantities.  One quantity to consider is the 
extrinsic curvature $\Theta_{MN}$ of the event horizon.
$\Theta_{MN}$ can be constructed from the null normal $n_M$
to the horizon and an auxiliary null vector $\ell_{M}$ whose normalization is 
conveniently chosen to satisfy $\ell_M n^M = -1$.  The extrinsic curvature is then given by 
$\Theta_{MN} \equiv \Pi^P_{\ M} \Pi^{Q}_{\ N} \nabla_P n_Q $
with $\Pi^M_{\ N} \equiv \delta^{M}_{\ N} + \ell^M n_N$.  
Since the horizon is at $r \approx 1$ we choose $n_M dx^M = dr$ and $\ell_{M} dx^M= -dt$.   
The horizon curvature satisfies $\Theta^M_{\ N} \Theta^N_{\ M} = \Theta^i_{\ j} \Theta^j_{\ i}$
where $i,j$ run over the spatial coordinates.   
For later connivence we define the rescaled traceless horizon curvature
$\theta^{i}_{\ j} \equiv \sqrt[4]{\frac{\gamma}{\kappa^2}} \Sigma^{i}_{ \ j}$ where 
$\Sigma^{i}_{\ j} \equiv  \Theta^i_{\ j} - \frac{1}{d} \Theta^{n}_{\ n} \delta^{i}_{\ j}$ is the traceless 
part of the extrinsic curvature and $\kappa$ is defined by the geodesic equation $n^M \nabla_M n_Q = \kappa n_Q$.

Also included in Fig.~\ref{fig:spectrum} are plots of $\mathcal A(t,k)/\mathcal P(t,k)$ where the horizon curvature power spectrum is
\begin{equation}
\label{eq:Adef}
\mathcal A(t,k) \equiv \frac{\partial}{\partial k} \int\limits_{|\bm k'| \leq k}  \frac{d^d k'}{(2 \pi)^d}  \tilde {\theta}^{*i}_{\ \ j}(t,\bm k') \tilde {\theta}^{j}_{\ i}(t,\bm k'),
\end{equation}
with 
$\tilde {\theta}^{i}_{\ j} \equiv \int d^d x \, \theta^i_{\ j}e^{-i \bm k \cdot \bm x}$.
As Fig.~\ref{fig:spectrum} makes clear, our numerical results are consistent with
\begin{equation}
\label{eq:AP}
\mathcal A(t,k) \sim k^2 \mathcal P(t,k).
\end{equation}
Evidently, bulk quantities --- even at the horizon ---  are correlated 
with boundary quantities.  As we detail below, this is a consequence 
of the applicability of the fluid/gravity correspondence.

Both qualitative and quantitative aspects of our results 
can be understood in terms of relativistic conformal hydrodynamics and the
fluid/gravity correspondence.  
In the limit of asymptotically slowly varying fields (compared to the dissipative scale set by the local temperature $T$ of the system)
Einstein's equations (\ref{eq:ein}) can be solved perturbatively with a gradient expansion 
$g_{\mu \nu}(x^\mu,r) = \sum_n g_{\mu \nu}^{(n)}(x^\mu,r)$ where $g_{\mu \nu}^{(n)}$ is order $\left (\partial/\partial x^\mu \right )^n$ in 
boundary spacetime derivatives \cite{Bhattacharyya:2008jc}.  
The expansion coefficients $g_{\mu \nu}^{(n)}$ can be expressed in terms of the boundary 
quantities $T$ and $u^\mu$ and their spacetime 
derivatives.  
The leading order term $g_{\mu \nu}^{(0)}$ is just the locally boosted black brane (\ref{eq:gradmetric}).
Likewise, via Eq.~(\ref{eq:bdstress}) the bulk gradient expansion 
encodes the boundary stress gradient expansion $\langle T_{\mu \nu}(x^\mu) \rangle = \sum_n T_{\mu \nu}^{(n)}(x^\mu)$.
The expansion coefficient  $T_{\mu \nu}^{(0)} = \frac{\epsilon}{d}  \, \left [ \eta_{\mu \nu} + (d+1) \, u_\mu u_\nu \right]$ is the stress tensor of ideal conformal hydrodynamics.
Likewise, $T_{\mu \nu}^{(1)} = - \eta\, \sigma_{\mu \nu}$ is the viscous stress tensor of conformal hydrodynamics with $\eta$ the shear viscosity and 
$\sigma_{\mu \nu}$ the shear tensor given below in (\ref{eq:shear}).
The underlying evolution of $u^\mu$ and $T$ and hence the bulk
geometry is governed by conservation of the boundary stress tensor. 
Hence at leading order in gradients the evolution of $u^\mu$ and $T$ is governed by ideal relativistic hydrodynamics
and the geometry is given by the boosted black brane metric (\ref{eq:gradmetric}).  Indeed, it was 
recently demonstrated in  \cite{Carrasco:2012nf} that turbulence in $d = 2$ ideal conformal relativistic hydrodynamics gives 
rise to an inverse cascade and exhibits the Kolmogorov scaling (\ref{eq:kol}).

\begin{figure}[ht!]
\includegraphics[scale = 0.25]{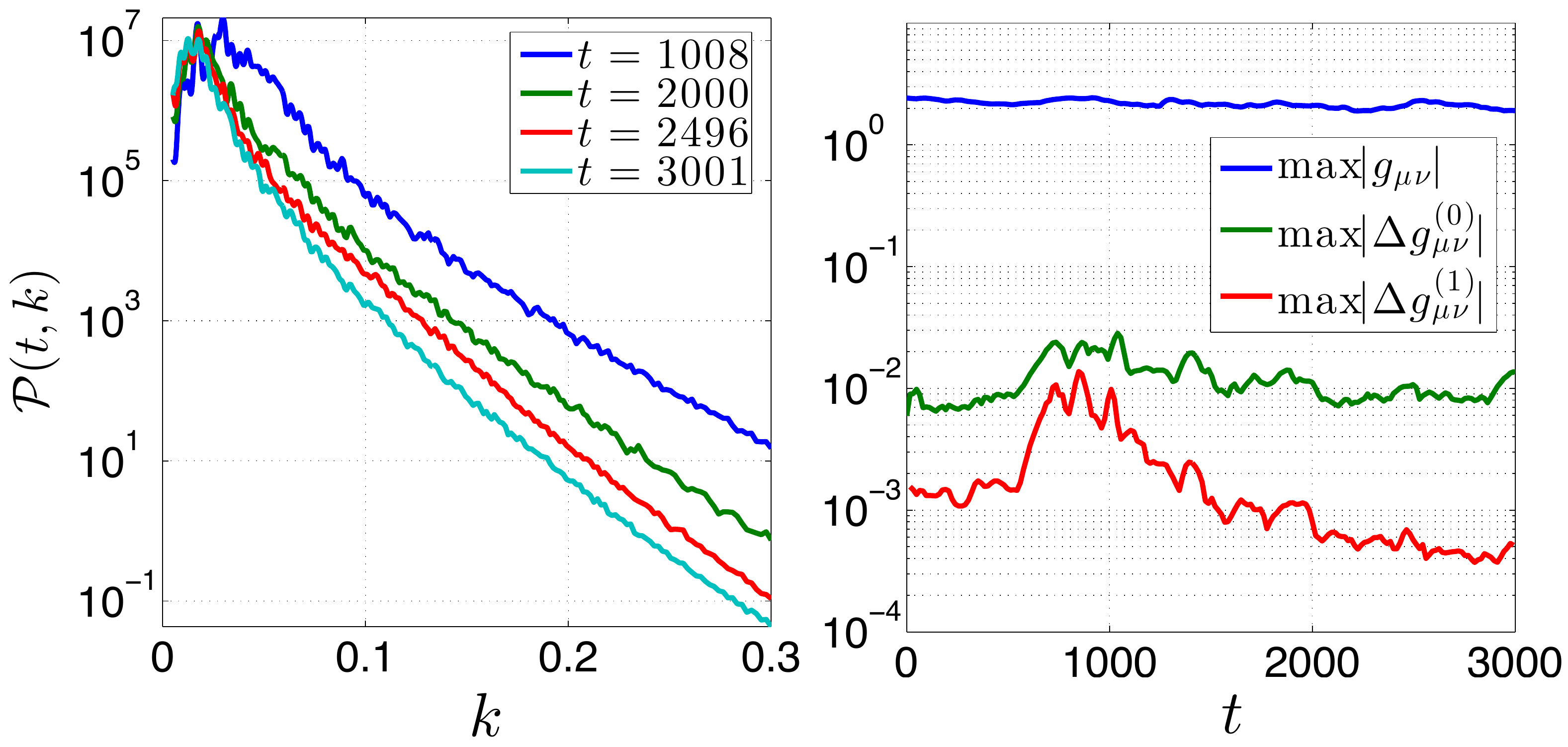}
\caption{Left: The velocity power spectrum on a semi-log scale at 4 different times.
Right: Time evolution of the maximum difference between the exact metric and zeroth and first order gradient expansion.
\label{fig:error}
} 
\end{figure}

Our first hint of the applicability of the fluid/gravity gradient expansion comes from
studying the power spectrum of the fluid velocity at high $k$ and the relative importance of viscous effects.  
In Fig.~\ref{fig:error} we plot $\mathcal P$ at for different times
on a semi-log scale.  As shown in the figure, 
our results are consistent with exponential decay of $\mathcal P$ 
for $k \gtrsim k_{\rm UV}$ with $k_{\rm UV} \sim 0.15$.  
With our choice of units where $4 \pi T/3 \sim 1$, the kinematic viscosity $\nu \equiv \frac{\eta}{\epsilon + p} \sim \frac{2}{3}$.
Therefore, at the scale $k_{\rm UV}$ viscous effects are suppressed by an order ten factor relative to inertial effects.
This suggest that the hydrodynamic gradient expansion on the boundary is well behaved.

Likewise, we find that our numerical metric $g_{\mu \nu}$
is  very well approximated by the fluid/gravity gradient expansion. 
To perform the comparison, via Eq.~(\ref{eq:fluidvel})  we extract $u^\mu$ and $\epsilon$ (and hence $T = (8 \pi G_{\rm N} \epsilon)^{1/3}$)
from  $\langle T^{\mu \nu} \rangle$. We then use $u^\mu$ and $T$ to construct the expansion functions $g_{\mu \nu}^{(n)}$
computed in \cite{VanRaamsdonk:2008fp}. 
We then take the difference $\Delta g_{\mu \nu}^{(N)} \equiv g_{\mu \nu} {-} \sum_{m = 0}^N g_{\mu \nu}^{(m)}$ and define the $N^{\rm th}$ order error to be
${\rm max} \{ | \Delta g_{\mu \nu}^{(N)}|  \}$ at 
each time $t$. As shown in Fig.~\ref{fig:error}, the boosted black brane metric (\ref{eq:gradmetric}) approximates the 
geometry at the 1\% level.  Including first order gradient corrections further decreases the size of the error 
\footnote
  {
  Near time $t = 1000$, when ${\rm max}| \Delta g_{\mu \nu}^{(0)}| \sim {\rm max}| \Delta g_{\mu \nu}^{(1)}|,$ 
  the size of the first order corrections to the metric
  is comparable to the truncation error of our numerical calculation.  In particular, by monitoring the 
  validity of temporal constraint equations we can ascertain the degree of convergence of our numerics.  Near time $t = 1000$
  the constraints are satisfied at order 1 part in $10^3$ which is the same size as the first order gradient correction 
  to the geometry.  The degree in which the constraints are satisfied can be improved by decreasing the spatial 
  grid spacing.  We suspect that using a finer resolution will decrease $ {\rm max}| \Delta g_{\mu \nu}^{(1)}|.$
  }.

It is natural that $d = 2$ turbulent evolution gives rise to dual geometries
well approximated by locally boosted black branes.  First of all, irrespective of $d$
turbulent flows require Reynolds number $Re \gg 1$, or equivalently, very small gradients
compared to $T$.  This is precisely the regime where the gradient expansions of fluid/gravity 
should be well behaved.  Second, the inverse cascade of $d = 2$  turbulence implies that  
gradients become smaller and smaller as energy cascades from the UV to the IR.
Therefore, the leading term (\ref{eq:gradmetric}) should become a better and better approximation to the metric as time progresses and the inverse cascade develops.

At least for $d = 2$ the above observation has powerful consequences for studying turbulent black holes.  
Instead of numerically solving the equations of general relativity one can simply study the equations
of hydrodynamics and construct the bulk geometry via the fluid/gravity gradient expansion.  
This is particularly illuminating in the limit of non-relativistic fluid velocities $|\bm u| \ll 1$, where the bulk geometry 
and boundary stress are asymptotically close to equilibrium.
As shown in \cite{Bhattacharyya:2008kq}, under the rescalings  
$t \to t/s^2,\ \bm x \to \bm x/s, \ \bm u \to s \bm u, \ \delta T \to s^2 \delta T,$
with $\delta T$ the variation in the temperature away from equilibrium, in the limit $s \to 0$ the boundary evolution of $\delta T$ and $u^\mu$ implied by the fluid/gravity correspondence
reduces to the non-relativistic incompressible Navier-Stokes equation.  Indeed, the above rescalings are 
symmetries of the Navier-Stokes equation.
Likewise, in the $s \to 0$ limit the geometry dual to the Navier-Stokes equation 
is encoded in the expansion functions $g_{\mu \nu}^{(0)}$ and $g_{\mu \nu}^{(1)}$ which are known analytically  \cite{Bhattacharyya:2008jc, VanRaamsdonk:2008fp}.
At least for $d = 2$, where is it well known that solutions to the Navier-Stokes equation are stable, 
we therefore expect many known results from classic studies of turbulence -- such as the Kolmogorov scaling (\ref{eq:kol}) -- to carry over naturally and semi-analytically to gravity.

As an illustration of the above point we now turn our attention to the horizon of turbulent black holes
and argue that the Kolmogorov scaling (\ref{eq:kol}) together with the relation (\ref{eq:AP})
implies that the turbulent horizons are fractal-like in nature with non-interger fractal dimension.  To augment our numerical evidence
of (\ref{eq:kol}) and (\ref{eq:AP}) we assume the validity of the fluid/gravity gradient expansion
for any $d$ and that the system is driven by an external force 
into a statistically steady-state configuration and that the Kolmogorov scaling applies over an arbitrarily large and static inertial range.
Within the gravitational description the driving can be accomplished by a time-dependent 
deformation of the boundary geometry \cite{Chesler:2008hg, Bhattacharyya:2008kq}.   


The fractal dimension of the horizon can be extracted from the 
horizon area.  Introducing a spatial regular $\delta x$, one could
compute the horizon area $A$ via the Riemann sum $A \approx \Sigma_i \sqrt{\gamma(x_i)} \Delta^d x_i$
where each element $ \Delta^d x_i \sim (\delta x)^d$.  The fractal dimension $D$ 
is  defined by the scaling
\begin{equation}
\label{eq:frataldimdef}
A \sim \left (\delta x \right )^{d - D},
\end{equation}
in the $ \delta x \to 0$ limit.   

To see how the Kolmogorov scaling (\ref{eq:kol}) implies the horizon has a fractal structure we employ the Raychaudhuri equation, 
\begin{equation}
\kappa\,\mathcal L_n \sqrt{\gamma} + {\zeta \over \sqrt{\gamma} } (\mathcal L_{n} \sqrt{\gamma})^2 - \mathcal L^2_n \sqrt{\gamma} = \sqrt{\gamma} \, \Sigma^{i}_{\ j} \Sigma^{j}_{\ i},
\end{equation}
which relates the change in the horizon's area element $\sqrt{\gamma}$ to the traceless part of horizon's extrinsic curvature $\Sigma^{i}_{\ j}$
\footnote 
  {
  For a review of this topic see~\cite{Gourgoulhon:2005ng}) 
  }.
Here $\mathcal L_n \equiv n^M \partial_M$ is the Lie derivative with 
respect to the null normal to the horizon $n_M$ and $\zeta=\frac{d{-}2}{d{-}1}$.
In the hydrodynamic limit of slowly varying fields salient to  turbulent flows the Raychaudhuri  equation simplifies to
$\mathcal L_n  \sqrt{\gamma}= \frac{\sqrt{\gamma}}{\kappa} \, \Sigma^{i}_{\ j} \Sigma^{j}_{\ i}.$
Integrating over the horizon, it follows that 
the rate of change of the horizon area $A$ is
\begin{equation}
\label{eq:horizonarea}
\frac{dA}{dt} = \int d^d x \frac{\sqrt{\gamma}}{\kappa} \, \Sigma^{i}_{\ j} \Sigma^{i}_{\ j} = \int_0^\infty dk \mathcal A(t,k),
\end{equation}
where $\mathcal A$ is given in (\ref{eq:Adef}).  We therefore see that 
$\mathcal A$ encodes the growth of the horizon area.

Using the locally boosted black brane metric (\ref{eq:gradmetric}), $\Sigma^{i}_{\ j}$ can 
be computed as a gradient expansion.  
For geometries dual to fluid flows in $d$ spatial dimensions the leading order results read \cite{Eling:2009sj}
\begin{equation}
\label{eq:extrinsichydrolimit}
{\textstyle \sqrt[4]{\frac{\gamma}{\kappa^2}} } \Sigma^{i}_{ \ j}= {\textstyle \frac{1}{\sqrt{2 \pi T}} \left ( \frac{ 4 \pi T}{d + 1} \right )^{d/2} } \left [ \sigma^{i}_{ \ j} + {\textstyle \frac{u^i}{u^0}} \sigma^{0}_{\ j} \right ]+ O(\partial^2),
\end{equation}
with $\sigma^{\mu}_{\ \nu} \equiv \eta^{\mu \alpha} \sigma_{\alpha \nu}$ and
\begin{equation}
\label{eq:shear}
\sigma_{\mu \nu } = \partial_{(\mu} u_{\nu)} + u_{(\mu} u^\rho \partial_\rho u_{\nu)} 
- {\textstyle \frac{1}{d}} \partial_\alpha u^\alpha \left [ \eta_{\mu \nu} {+}u_\mu u_\nu \right].
\end{equation}
We note that $\sigma_{\mu \nu}$ is orthogonal to the fluid velocity $u^\mu \sigma_{\mu \nu} = 0$ and traceless $\eta^{\mu \nu} \sigma_{\mu \nu} = 0$.
Using (\ref{eq:Adef}) and (\ref{eq:extrinsichydrolimit}) we see that at leading order in gradients $\mathcal A(t,k)$ is the 
power spectrum of 
$(2 \pi T)^{-1/2}\left [4 \pi T/(d {+} 1)\right ]^{d/2} \sigma^\mu_{\ \nu}$.  
Counting derivatives we therefore see that (\ref{eq:AP}) must be satisfied at leading 
order in gradients, just as demonstrated in Fig.~\ref{fig:spectrum}.

Assuming that the system is driven into a steady-state and the Kolmogorov scaling (\ref{eq:kol}) applies,
from (\ref{eq:AP}) we see that for any $d$ we have
$\mathcal A \sim k^{1/3}$.
Inserting a UV regulator into the momentum integral in (\ref{eq:horizonarea}) at $k = k_{\rm max}$ we conclude that for $k_{\rm max} \in (\Lambda_-,\Lambda_+)$
\begin{equation}
\label{eq:areascaling}
d A(k_{\rm max})/dt \sim k_{\rm max}^{4/3} \ .
\end{equation}
Integrating and identifying $\delta x \sim 1/k_{\rm max}$ we conclude from (\ref{eq:frataldimdef})
that the fractal dimension of the horizon is
\begin{equation}
\label{eq:fractaldim}
D = d + 4/3.
\end{equation}

We note, however,
that the scaling  (\ref{eq:frataldimdef}) is never 
exactly obtained for a turbulent horizon.  
The UV terminus $\Lambda_+$ of the inertial range is bounded by the dissipative scale $T$,
where hydrodynamics
breaks down.  Hence the scaling (\ref{eq:areascaling}) can apply no further than $k_{\rm max} \lesssim T$.
However, the constant of integration one gets from integrating (\ref{eq:areascaling})
is of order $T^d$, which is the area element of an equilibrium black brane.
Hence the $k_{\rm max}^{4/3}$ scaling of the horizon area never 
dominates over the $T^d$ constant and the scaling  (\ref{eq:frataldimdef}) is never 
exactly satisfied in the large $k_{\rm max}$ limit.  However, the domain $k_{\rm max} \in (\Lambda_-,\Lambda_+)$ 
in which the scaling (\ref{eq:areascaling}) 
applies can be made arbitrarily large as $\Lambda_-$ is bounded only by the system size.
Therefore, what Eqs.~(\ref{eq:areascaling}) and (\ref{eq:fractaldim}) encode
is that turbulent horizons have geometric features over a large hierarchy of scales,
just as seen in the plots of $\sqrt{\gamma}$ in Fig.~\ref{fig:vortictyandarea}.

The origin of the fractal-like structure of the horizon is easy to understand.  It is well known from fluid mechanics that
turbulent flows have a fractal-like structure with large vortices being composed of smaller vortices which are themselves composed 
of smaller vortices and so on.  This behavior can be seen in the plots of the 
vorticity in Fig.~\ref{fig:vortictyandarea}.
Via the fluid/gravity gradient expansions this fractal-like structure imprints itself on the bulk geometry.
Moreover, upon using Hawking's formula to relate the horizon area to entropy, the Raychaudhuri equation
(\ref{eq:horizonarea}) combined with (\ref{eq:extrinsichydrolimit}) translates into the familiar expression for entropy growth in hydrodynamics
$dS/dt = \int d^d x \, \frac{2 \eta}{T} \, \sigma_{\mu \nu} \sigma^{\mu \nu}$ with $\eta$ the shear viscosity of the 
fluid (which for theories with gravitational duals is given by $\eta = s/4 \pi$ with $s$ the proper entropy density \cite{Kovtun:2004de}) \cite{Eling:2009sj}.
Thus the fractal horizon can be understood as the geometric counterpart of the 
familiar fact that turbulent flows generate entropy much more rapidly than laminar flows with the same rate of energy dissipation.

We conclude by discussing the origin of the inverse cascade for $d = 2$.
It is well known that  the $d = 2$  inviscid Navier-Stokes equation conserves the 
enstrophy $\int d^2 x\, \omega^2$ and that conservation of enstrophy necessitates and inverse cascade. 
Moreover, it has recently been demonstrated that 
the equations of $d = 2$ ideal relativistic conformal hydrodynamics salient to holographic turbulence
conserve the relativistic generalization of the enstrophy  
\begin{equation}
\label{eq:relativisticenstrophy}
\Omega \equiv \int d^2 x \, \omega^2 u^0,
\end{equation}
\cite{Carrasco:2012nf}.
Conservation of $\Omega$ in $d = 2$ hydrodynamics suggests that there should exist an analogous conserved quantity in the 
gravitational description which is also only conserved for $d = 2$.   Since the enstrophy is constructed from the vorticity $\omega$, a natural
starting point is finding a gravitational quantity which encodes $\omega$.
As $\omega$ is a pseudoscalar in two spatial dimensions a natural guess is the gravitational Pontryagin density 
$^{*}RR \equiv  \frac{1}{\sqrt{-g}} \epsilon^{MNAB} R^{Q}_{ \ \ PAB} R^{P}_{ \ \ QMN}$.  
With this guess we define the gravitational
enstrophy 
\footnote
  {
  We note there are other gravitational quantities that encode both the 
  vorticity and enstrophy.
  }
\begin{equation}
\Omega_{\rm grav} \equiv \int d^2 x \sqrt{\gamma} \, ( ^{*}RR )^2,
\end{equation}
where the integration is over the horizon.  
Using the boosted black brane metric (\ref{eq:gradmetric}) and expanding in powers of derivatives we find 
$\sqrt{\gamma} \, (^{*}RR)^2 = (72 \, \omega)^2 u^0 + O(\partial^2).$
Hence up to a numerical factor $\Omega_{\rm grav}$ 
coincides with the fluid enstrophy $\Omega$.  Evidently, in the long
wavelength limit evolution generated by Einstein's equations must 
conserved $\Omega_{\rm grav}$.

\acknowledgments

\textit{Acknowledgments.}---We acknowledge helpful conversations with Christopher Herzog, Veronika Hubeny, Luis Lehner, Mukund Rangamani, and Laurence Yaffe.  
AA thanks the Aspen Center for Physics for hospitality.
The work of AA is supported in part by the U.S. Department of Energy (D.O.E.) under cooperative research agreement \#DE-FC02- 94ER40818.
The work of PC is supported by a Pappalardo Fellowship in Physics at MIT.  
The work of HL is partially supported by a Simons Fellowship and by the U.S. Department of Energy (D.O.E.) under cooperative research agreement \#DE-FG0205ER41360.

\bibliography{refs}%
\end{document}